\documentclass [12pt,a4paper,leqno]{article}

\def\R{\mathrm{R}}

\title{A Stationary Asymptotically Einstein Static Universe Solution}
\author{Mihaela Time}
\date{7th~September~1998}

\begin{document}

\maketitle

\begin{abstract}
Here I present a stationary cylindrically symmetric asymptotically Einstein static
universe solution with the matter consisting of a cosmological and rotating dust
term which admits predicted black hole event horizon.

\end{abstract}

\section{Introduction}
I will consider here the strong gravitational fields in some form of
gravitational collapse, when the gravitational forces cannot be compensated by
other forces. As a special case I introduce an asymptotically Einstein static
universe space-time. This space-time is far from the strong gravitational field
and the metric has the form of the Einstein static universe at infinity.

The spatially closed, static Einstein universe,
\begin{eqnarray} \label{eq:E}
\mathrm{d}s_E^2=\mathrm{d}\eta ^2+\sin ^2\eta(\mathrm{d}\theta ^2+\sin ^2\theta
\mathrm{d}\varphi ^2)-c^2\mathrm{d}\psi ^2\\
\varphi\in[0,2\pi ],\quad \eta\in [0,\pi ],\quad \theta\in [0,\pi ],
\quad \psi \in \R \nonumber
\end{eqnarray}
is the simplest cosmological dust model with constant curvature $K=\mathrm{const.}$;
the energy-momentum tensor $T_{ab}$ has the form (\ref{eq:momentum}), where
$\Lambda={1\over K^2}$ and $\mu={2\over K^2}=2\Lambda$.

The space-time of Special Relativity is described mathematicaly by the Minkowski
space $(M,\, \eta)$. The flat metric $\eta$, in coordinates \footnote{The metric
is apparently singular for $r = 0$ and $\sin \theta = 0$.}
\begin{eqnarray} \label{eq:M}
\mathrm{d}s^2=\mathrm{d}r^2+r^2(\mathrm{d}\theta ^2 +\sin^2\theta
\mathrm{d}\varphi ^2 ) - c^2\mathrm{d}t^2 \\
r\in [0, \infty), \quad \theta \in [0, \pi], \quad \varphi \in [0, 2\pi], \quad
t \in \R, \nonumber
\end{eqnarray}
is wholly conformal to a finite region of the Einstein static universe
(\ref{eq:E}).

We have:
\begin{equation} \label{eq:tr}
 \mathrm{d}s^2 = \Theta^{-2}\mathrm{d}s_E^2,
\end{equation}
where $\Theta = 2\cos\frac{\psi+\eta}{2}\cos\frac{\psi-\eta}{2}$ is a smooth
strictly positive function, under the conformal transformation:
$$t+r = \mathrm{tg}\frac{\psi+\eta}{2}, \enspace t-r = \mathrm{tg}\frac{\psi-
\eta}{2}, \enspace -\frac{\pi}{2}\leq \psi-\eta \leq \psi+\eta \leq
\frac{\pi}{2}$$
(the boundaries $\psi\pm\eta = \pm\frac{\pi}{2}$ are the null surfaces
$\mathcal{I}^+$ and $\mathcal{I}^-$).

The de Sitter space-times are also conformal to a (finite) part of
$\mathrm{d}s_E^2$ and generally, all the closed Robertson-Walker metrics
(Minkowski space, the Sitter space are included as special cases) are
conformally equivalent to the Einstein static universe.
The Robertson-Walker metric,
\begin{equation}
\mathrm{d}s_{R-W}^2=K^2(ct)[\mathrm{d}\eta ^2+\sin ^2\eta (\mathrm{d}\theta^2 +
\sin ^2\theta \mathrm{d}\varphi ^2)]-c^2\mathrm{d}T^2
\end{equation}
under the transformation
\begin{equation} \label{eq:tbeg}
 \mathrm{d}T = \frac{1}{K(ct)}\mathrm{d}\psi
\end{equation}
gives (\ref{eq:E}) .

In this paper I will construct a solution of the Einstein dust equation
$(M, g)$, which asimptotically approaches the Einstein static universe. This
asimptotic limit requires the energy-momentum tensor to have the form of dust
and $\Lambda \neq 0$ to be positive.

Here I shall adopt in the definition of this kind of asymptotically Einstein
static universe space-times that the boundary $\partial M$ is consisted by two
disjoint (closed) spacelike \footnote{In asymptotically flat spaces,
$\mathcal{I}^+$ and $\mathcal{I}^-$ are null surfaces each of which of
$\R^{1}\times S^2 $ topology.} surfaces $\mathcal{I}^+$ and $\mathcal{I}^-$(for
the Robertson-Walker metrics,
$\mathcal{I}^+$ and $\mathcal{I}^-$ are $T_b$, respectively $T_e$, the beginning
of world and the end of universe).

\section{A stationary asymptotically Einstein static universe solution}
Stationary gravitational fields are characterized by the existence of a timelike
Killing vector field $\xi$. So, in a stationary space-time $(M,\,g)$ we can have
a global causal structure, i.e. we can introduce an adapted coordinate system
$(x^a) = (x^{\alpha},\,t)$, $\xi = \frac{\partial}{\partial t}$ so that the
metric $g_{ab}$ is independent of $t$,
\begin{equation} \label{eq:stat}
 \mathrm{d}s^2 = h_{\alpha\beta}\mathrm{d}x^{\alpha}\mathrm{d}x^{\beta} +
F{(\mathrm{d}t + A_{\alpha}\mathrm{d}x^{\alpha})}^2, \qquad F \equiv \xi_a\xi^a
< 0.
\end{equation}
The timelike unitary vector field $h^0 \equiv {(-F)}^{-{1\over 2}}\xi$, which is
globally defined on $M$, indicates not only the time-orientation in every point
$p \in M$, but also gives a global time coordinate $t$ on $M$.

Stationarity (i.e. time translation symmetry) means that there exists a
1-dimensional group $G_1$ of isometries $\phi_t$ whose orbits are timelike
curves parametrized by $t$.

Using the 3-projection formalisme (Geroch (1971)) developed on the 3-dimensional
differentiable factor manifold $\mathcal{S}_3$, the Einstein's
field equations
\begin{equation}
 R_{ab} - {1\over 2} R g_{ab} + \Lambda g_{ab} = \kappa T_{ab},
\end{equation}
for stationary fields take the following simplified form:
\begin{equation} \label{eq:einstein}
\left \{
\begin{array}{l}
\displaystyle R^{(3)}_{ab}={1\over 2}F^{-2}(\frac{\partial F}{\partial
x^a}\frac{\partial F}{\partial x^b}+\omega_{a}\omega_{b})
+\kappa(h^c_ah^d_b-F^{-2}\tilde{h}_{ab}\xi^a\xi^b)(T_{cd}-{1\over 2}Tg_{cd});

\vspace{6pt} \\ \vspace{6pt}

\displaystyle F^{\parallel a}_{,a}= F^{-1}\tilde{h}_{ab}(\frac{\partial
F}{\partial x^a}\frac{\partial F}{\partial x^b}-\omega_{a}
\omega_{b})-2\kappa F^{-1}\xi^a\xi^b(T_{ab}-{1\over 2}Tg_{ab});\\
\displaystyle \omega^{\parallel a}_a=2F^{-1}\tilde{h}_{ab}\frac{\partial
F}{\partial x^a}\omega_b

\vspace{6pt} \\ \vspace{6pt}

\displaystyle F \epsilon^{abc} \omega_{c,b} = 2 \kappa h_b^a T_c^b \xi^c
\end{array}
\right .
\end{equation}

Here, ``$\parallel$'' denotes the covariant derivative associated with the
conformal metric tensor $\tilde{h}_{ab} = -F h_{ab}$ on $\mathcal{S}_3$
($h_{ab} = g_{ab} + h^0_a h^0_b$ is the projection tensor) and $\omega^a =
{1\over 2}\epsilon^{abcd} \xi_{b} \xi_{c;d} \neq 0$ \footnote{Here I use the
convention: round brackets denote symmetrization and square brackets
antisymmetrization and $\Omega$ is the angular velocity.} is the rotational vector
($\omega^a\xi_a=0,\enspace \pounds_\xi\omega=0$).

This Einstein static universe asymptotic limit suggests there are more than one
symmetry on $M$. I suppose the metric $g_{ab}$ admits also an Abelian group of
isometries $G_2$ generated by the two spacelike Killing vector fields $\eta$ and
$\zeta$, $\pounds_{\eta}g_{ab} = \pounds_{\zeta}g_{ab} = 0, \enspace
\eta_a\eta^a > 0, \enspace \zeta_a\zeta^a > 0$ and the integral curves of
$\eta$ are closed (spatial) curves.

The circularity theorem (due to Kundt) states that an (axisymmetric) metric can
be written in the (2+2)-split if and only if the conditions
\begin{equation}
(\eta^{[a}\xi^b\xi^{c;d]})_{;e}=0=(\xi^{[a}\eta^b\eta^{c;d]})_{;e}
\end{equation}
are satisfied.

The existence of the orthogonal 2-surfaces is possible for dust solutions,
provided that the 4-velocity of dust satisfies the condition:
\begin{eqnarray}
u_{[a}\xi_b\eta_{c]}=0,\quad u^a=(-H)^{-{1\over 2}}(\xi^a+\Omega\eta^a)=
(-H)^{-{1\over 2}}l^i\xi^a_i,\quad\textrm{where}\\
l^i\equiv(1,\Omega),\quad H=\gamma_{ij}l^il^j,
\quad \gamma_{ij}\equiv \xi^a_i\xi_{aj},\quad i,j=1,2,\quad \xi_1=\xi;\ \xi_2=\eta \nonumber
\end{eqnarray}
)
i.e., the trajectories of the dust lie on the transitivity surfaces of the group
generated by the Killing vectors $\xi$, $\eta$.

Using an adapted coordinate system, the metric (\ref{eq:stat}) can be written in
the following form:
\begin{equation} \label{eq:dust}
\mathrm{d}s^2=e^{-2U}[e^{2V}(\mathrm{d}r
^2+\mathrm{d}z^2)+W^2\mathrm{d}\varphi ^2]-
e^{2U}(\mathrm{d}t+A\mathrm{d}\varphi )^2
\end{equation}
where the functions \footnote{The function $W$ is defined invariantly as
$W^2 \equiv -2\xi_{[a}\eta_{b]}\xi^a\eta^b$.} $U$, $V$, $W$ and $A$
depend only on the coordinates $(r,z)$; these coordinates are also conformal
flat coordinates on the 2-surface $S_2$ orthogonal to 2-surface $T_2$ of the
commuting Killing vectors $\xi = \partial_t$ and $\eta = \partial_\varphi$.

If we identify the 4-velocity of the dust $u^a$ with timelike Killing vector
$\xi^a = \partial_t = (0, 0, 0, 1)$ then (\ref{eq:dust}) represents a co-moving
system $(x^1=r,\ x^2=z,\ x^3=\varphi,\ x^0=t)$ with dust,
$u_a = \xi_a = (0,\, 0,\, -e^{2U}A,\, -e^{2U})$ and
\begin{equation} \label{eq:metric}
\left \{
\begin{array}{l}
g_{11}=g_{22}=e^{-2U+2V}=h_{11}=h_{22},\\
g_{33}=e^{-2U}W^2-e^{2V}A^{2}=h_{33},\quad g_{00}=\xi_0=-e^{2U}=F,\\
g_{03}=\xi_3=-e^{2U}A,\quad g_{13}=g_{23}=g_{10}=g_{20}=0
\end{array}
\right .
\end{equation}

We can use the complex coordinates $(q, \bar{q})$ on the 2-surface $S_2$:
\begin{equation}
 q={1\over \sqrt{2}}(r+iz)
\end{equation}

The surface element on $T_2$ is $f_{ab}=2\xi_{[a}\eta_{b]}$, $f_{ab}f^{ab}<0$
and the surface element on $S_2$ is
$\tilde{f}_{ab}$, the dual tensor of $f_{ab}$,
$\tilde{f}_{ab}={1\over 2}\epsilon_{abcd}f^{cd}$

Thus the Einstein's dust equations with constant $\Lambda > 0$
(\ref{eq:einstein}) for the metric (\ref{eq:metric}) will take the following
form:
\begin{equation}\label{eq:einstein2}
\left \{
\begin{array}{l}
\displaystyle
\frac{\partial^{2}W}{\partial q\partial \bar{q}}=-\Lambda We^{2V-2U}
\vspace{6pt} \\ \vspace{6pt}
\displaystyle
\frac{\partial^{2}U}{\partial q\partial \bar{q}}+\frac{1}{2W}(\frac{\partial
U}{\partial q} \frac{\partial W}{\partial \bar{q}}
+\frac{\partial U}{\partial \bar{q}} \frac{\partial W}{\partial
q})+\frac{1}{2W^2}e^{4U} \frac{\partial A}{\partial q}
\frac{\partial A}{\partial \bar{q}} = (\mu-2\Lambda)\frac{e^{2V-2U}}{4}
\vspace{6pt} \\ \vspace{6pt}
%\noalign{\medskip}
\displaystyle
\frac{\partial^{2}A}{\partial q\partial \bar{q}}-\frac{1}{2W}(\frac{\partial
A}{\partial q} \frac{\partial W}{\partial \bar{q}}
+\frac{\partial A}{\partial \bar{q}} \frac{\partial W}{\partial q}) +
2(\frac{\partial A}{\partial q} \frac{\partial U}{\partial \bar{q}}
+\frac{\partial A}{\partial \bar{q}} \frac{\partial U}{\partial q})=0
\vspace{6pt} \\ \vspace{6pt}
\displaystyle
\frac{\partial^{2}W}{\partial q\partial \bar{q}}-2\frac{\partial W}{\partial q}
\frac{\partial V}{\partial q}+
2W(\frac{\partial U}{\partial q})^2-\frac{1}{2W}e^{4U}(\frac{\partial
A}{\partial q})^2=0
\vspace{6pt} \\ \vspace{6pt}
\displaystyle
\frac{\partial^{2}V}{\partial q\partial \bar{q}}+\frac{\partial U}{\partial q}
\frac{\partial U}{\partial \bar{q}}
+\frac{1}{(2W)^2}e^{4U}\frac{\partial A}{\partial q} \frac{\partial A}{\partial
\bar{q}}=
 -\Lambda \frac{e^{2V-2U}}{2}
\end{array}
\right .
\end{equation}
Here $\displaystyle\Delta=\frac{\partial^2}{\partial r^2} +
\frac{\partial^2}{\partial z^2}=2\frac{\partial^2}{\partial q \partial\bar{q}}$
is the Laplace operator and the energy-momentum tensor $T_{ab}$ has the form as:
\begin{equation}\label{eq:momentum}
\kappa T_{ab}=-\Lambda g_{ab}+\mu u_a u_b,\quad \mu>0,
\Lambda =\textrm{const.}>0.
\end{equation}

The conservation law, $T_{;b}^{ab}=0$ implies $U_{,a}=0$ and is a consequence
of the field equations, being used in place of one of the equations (\ref{eq:einstein2}).

Using the simplification $U=0$ in (\ref{eq:metric}) then the matter current is geodesic
($\dot{u}_a=u_{a;b}u^b=0$) and without expansion ($\theta=u_{;a}^a=0$), but
has a non-rigidly rotation, ($\omega=\sqrt{{1\over 2}\omega_{ab}\omega^{ab}} \neq 0$)
and $\sigma \neq 0$; the field equations (\ref{eq:einstein2}) will take the following
simplified form:
\begin{equation} \label{eq:einstein3}
\left \{
\begin{array}{l}
\displaystyle
\frac{\partial^{2}W}{\partial q\partial \bar{q}}=-\Lambda We^{2V}

\vspace{6pt} \\ \vspace{6pt}

\displaystyle
\frac{1}{2W^2} \frac{\partial A}{\partial q} \frac{\partial A}{\partial \bar{q}}
 = (\mu-2\Lambda)\frac{e^{2V}}{4}

\vspace{6pt} \\ \vspace{6pt}

\displaystyle
\frac{\partial^{2}A}{\partial q\partial \bar{q}}-\frac{1}{2W}(\frac{\partial
A}{\partial q} \frac{\partial W}{\partial \bar{q}}
+\frac{\partial A}{\partial \bar{q}} \frac{\partial W}{\partial q}) = 0

\vspace{6pt} \\ \vspace{6pt}

\displaystyle
\frac{\partial^{2}W}{\partial q\partial \bar{q}}-2\frac{\partial W}{\partial q}
\frac{\partial V}{\partial q}-
\frac{1}{2W}(\frac{\partial A}{\partial q})^2=0

\vspace{6pt} \\ \vspace{6pt}

\displaystyle
\frac{\partial^{2}V}{\partial q\partial \bar{q}}
+\frac{1}{(2W)^2}\frac{\partial A}{\partial q} \frac{\partial A}{\partial
\bar{q}}= -\Lambda \frac{e^{2V}}{2}
\end{array}
\right .
\end{equation}

Finally, if we take into account the third symmetry ($\zeta = \partial_z$ the
spacelike Killing vector field) as a special case of the stationary
axisymmetric (with $\xi$ and $\eta$) solution (\ref{eq:einstein3}) and if we
match the parameters by a particular choice such that the exterior field is
conformal with Einstein static universe \footnote{For the case when $\Lambda =
0$ the extern field becomes static
($\omega^a\!=\!{1\over 2}\epsilon^{abcd}\xi_b\xi_{c;d}\!=0$) even the dust was
in rotation with $\Omega = \textrm{const.}$}, then we obtain a stationary
asymptotically Einstein static universe solution:
$$\mathrm{d}s^2=e^{2V(r)}(\mathrm{d}r^2 + \mathrm{d}z^2) +
W^2(r)\mathrm{d}\varphi - (\mathrm{d}t+A(r)\mathrm{d}\varphi)^2$$
where $V(r)$, $W(r)$, $A(r)$ and $\mu(r)$ depend only on $r$.

Indeed denoting ${\partial\over \partial r}='$ the field equations (\ref{eq:einstein3})
take the form:
\begin{equation} \label{eq:einstein4}
\left \{
\begin{array}{l}
\displaystyle W''=-2\Lambda We^{2V} \\
\displaystyle {A'}^2=(\mu-2\Lambda)W^2e^{2V} \\
\displaystyle \frac{A''}{A'} = \frac{W'}{W} \\
\displaystyle 2W''-4W'V'-{1\over W}A'^2=0 \\
\displaystyle V''+{1\over 4W^2}{A'}^2=-\Lambda^{2V}
\end{array}
\right .
\end{equation}

The third equation of the previous system is integrable: $A'=\alpha W$,
where $\alpha=\mathrm{const.} \neq 0$. Taking $\alpha=2$ and inserting
in the rest of the system we get:
\begin{equation} \label{eq:einstein5}
\left \{
\begin{array}{l}
\displaystyle \mu=2\Lambda +4e^{-2V} \\
\displaystyle \frac{W'}{W} = -\frac{1+\Lambda e^{2V}}{V'} \\
\displaystyle V''+\Lambda e^{2V}+1=0
\end{array}
\right .
\end{equation}

Because the cosmological constant $\Lambda$ is very small (less than
$10^{-57} cm^{-2}$) we can approximate $\Lambda e^{2V} \approx
\Lambda(1+2V)$. In this approximation the system gets the form:
\begin{equation} \label{eq:einstein6}
\left \{
\begin{array}{l}
\displaystyle \mu=2\Lambda +\frac{4}{1+2V} \\
\displaystyle \frac{W'}{W} = -\frac{1+\Lambda (1+2V)}{V'} \\
\displaystyle V''+\Lambda (1+2V)+1=0
\end{array}
\right .
\end{equation}

Integrating first the last linear equation ($V''+2\Lambda V+\Lambda+1=0$) we obtain:
\begin{equation}
\begin{array}{l}
\displaystyle 
V(r)=C_1\sin(\sqrt{2\Lambda}r+\alpha)-\frac{1+\Lambda}{2\Lambda} \\ \vspace{6pt}
\displaystyle
\mu(r)=2\Lambda+4\frac{1}{1+2C_1\sin(\sqrt{2\Lambda}r+\alpha)-\frac{1+\Lambda}{\Lambda}} \\
\vspace{6pt}
W(r)={1\over 2}C_3C_1\sqrt{2\Lambda}\cos(\sqrt{2\Lambda}r+\alpha) \\
\vspace{6pt}
\displaystyle
A(r)=C_3(C_1\sin(\sqrt{2\Lambda}r+\alpha)-\frac{1+\Lambda}{2\Lambda})+C_4
\end{array}
\end{equation}
where $\alpha, C_1, C_2, C_3$ are parameters of integration of the system which must be
chosen so that for $r\to \infty$ we have $\mu(r)\to 2\Lambda$ and
$A(r)\to 0$, so the solution found is asymptotically Einstein static universe solution...

\section{Some discussion}
Another type of event horizon, caused by the presence of the repulsive $\Lambda$
term, is defined as the boundary of the points of spacetime from which light can
never reach the observer. This boundary is called {\em the cosmological event
horizon} of the observer (see \cite{gb:cosmological}).

On such a space-time \footnote{No strong causality condition but only a stable
causal condition is holded on space-time.} which admits predicted black hole
event horizons and it is conformal with  Einstein static universe at infinity,
there are points from where light rays (null geodesics) that escape to null
infinity ${\mathcal I}^+$ would return to ${\mathcal I}^-$ on a proper finite
time interval (as in general closed universes) and would also return to their
source. Such a region of points
$p\in M, p\in J^+({\mathcal I}^-)\cap J^-({\mathcal I}^+)$
contains causal closed curves (null or temporal picewise geodesics).
Indeed, this is the case since the behaviour near infinity could be determined
by the light ray structure of space-time and that light rays remain light rays
(and causal structure has not changed either) of the Einstein static universe
under a conformal transformation ($\mathrm{d}s^2 = 0$ implies 
$\mathrm{d}s_E^2 = \Theta^2 \mathrm{d}s^2 = 0$)
and light rays of the Einstein universe have the property that they all return
to their source at a time interval to cover $\Delta T=\frac{2\pi}{c}$.

\end{document}